\documentclass[prl,twocolumn,superscriptaddress,noshowpacs,notitlepage,longbibliography]{revtex4-1}%
\usepackage{graphicx,bm,times}
\usepackage{amsmath}
\usepackage{amsfonts}
\usepackage{amssymb}
\usepackage{color}
\usepackage{xcolor}
\usepackage{hyperref}
\usepackage{textcomp}
\hypersetup{
	colorlinks = true,
	allcolors = {blue}
}

\begin{document}

\title{Strong surface termination dependence of the electronic structure of polar superconductor LaFeAsO revealed by nano-ARPES}

\author{Sung Won Jung}
\email[corresponding author: ]{sungwon.jung@gnu.ac.kr}
\affiliation{Diamond Light Source, Harwell Campus, Didcot, OX11 0DE, United Kingdom}
\affiliation{Department of Physics and Research Institute of Natural Science, Gyeongsang National University, Jinju 52828, Republic of Korea}
\author{Luke C. Rhodes}
\affiliation{School of Physics and Astronomy, University of St. Andrews, St. Andrews KY16 9SS, United Kingdom}	

\author{Matthew D. Watson}
\affiliation{Diamond Light Source, Harwell Campus, Didcot, OX11 0DE, United Kingdom}

\author{Daniil V. Evtushinsky}
\affiliation{Laboratory for Quantum Magnetism, Institute of Physics, \'{E}cole Polytechnique F\'{e}d\'{e}rale de Lausanne, CH-1015 Lausanne, Switzerland}
	
\author{Cephise Cacho}
\affiliation{Diamond Light Source, Harwell Campus, Didcot, OX11 0DE, United Kingdom}

\author{Saicharan Aswartham}
\affiliation{Leibniz Institute for Solid State and Materials Research, 01171 Dresden, Germany}

\author{Rhea Kappenberger}
\affiliation{Leibniz Institute for Solid State and Materials Research, 01171 Dresden, Germany}

\author{Sabine Wurmehl}
\affiliation{Leibniz Institute for Solid State and Materials Research, 01171 Dresden, Germany}

\author{Bernd B\"{u}chner}
\affiliation{Leibniz Institute for Solid State and Materials Research, 01171 Dresden, Germany}
\affiliation{Institut für Festkörperphysik, Technische Universit\"{a}t Dresden, D-01171 Dresden, Germany}

\author{Timur K. Kim}
\email[corresponding author: ]{timur.kim@diamond.ac.uk}
\affiliation{Diamond Light Source, Harwell Campus, Didcot, OX11 0DE, United Kingdom}

\begin{abstract}
The electronic structures of the iron-based superconductors have been intensively studied by using angle-resolved photoemission spectroscopy (ARPES). 
A considerable amount of research has been focused on the LaFeAsO family, showing the highest transition temperatures, where previous ARPES studies have found much larger Fermi surfaces than bulk theoretical calculations would predict. 
The discrepancy has been attributed to the presence of termination-dependent surface states. 
Here, using photoemission spectroscopy with a sub-micron focused beam spot (nano-ARPES) we have successfully measured the electronic structures of both the LaO and FeAs terminations in LaFeAsO. 
Our data reveal very different band dispersions and core-level spectra for different surface terminations, showing that previous macro-focus ARPES measurements were incomplete.
Our results give direct evidence for the surface-driven electronic structure reconstruction in LaFeAsO, including formation of the termination-dependent surface states at the Fermi level. 
This new experimental technique, which we have shown to be very powerful when applied to this prototypical compound, can now be used to study various materials with different surface terminations.
\end{abstract}
\date{\today}

\maketitle

\section{Introduction}
LaFeAsO is the parent compound of the so-called ``1111” family of iron-based superconductors – the family in which superconductivity was first discovered by Kamihara \textit{et al.}~\cite{Kamihara} and in which the highest reported superconducting transition temperatures have been found~\cite{Ren_2008, Wang_2008}. Despite being among the first compounds discovered, many questions remain as to its experimental electronic structure. This is partly because of the lower quality of the original single crystals, but more fundamentally is due to the fact that its structure includes alternately charged LaO and FeAs planes. Thus upon cleaving the sample, it can have two surface terminations, each of which is a polar surface with uncompensated charge, and unrepresentative of the bulk.  

\begin{figure}[t]
\centering
\includegraphics[width=\linewidth]{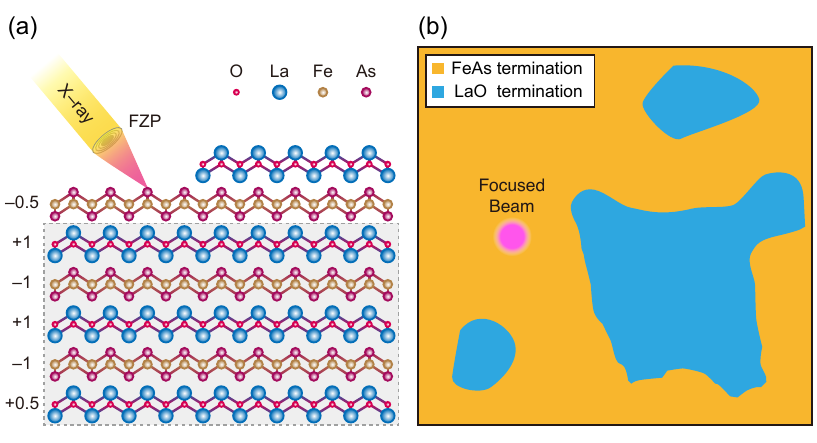}
\caption{\textbf{Termination dependent structure of LaFeAsO surface}
(a) layered crystal structure with polarised LaO and FeAs surface terminations;
(b) sketch of the cleaved crystal surface with various terminations.}
\label{fig1}
\end{figure}

Earlier angular resolved photoemission spectroscopy (ARPES) studies \cite{Shen_LaOFeAs,KaminskiLaFeAsO,Feng_LaOFeAs,Ding_LaOFeAs,CKim2020_LaFeAsO} revealed an electronic structure that substantially differed from the expectations from naive bulk density functional theory (DFT) calculations. In particular, the spectra included a bright hole-like band around the Brillouin zone center. 
It was quickly appreciated that this band, and several other prominent bands, were attributable to the surface FeAs layer which has an effective hole-doping~\cite{KaminskiLaFeAsO}, with the overall spectra being somewhat reminiscent of the hole-doped end-compound KFe$_2$As$_2$ but totally unlike the spectra on other nominally charge-compensated parent compounds such as BaFe$_2$As$_2$~\cite{Fedorov2019PRB_BaFe2As2,Matt_NanoARPES} and NaFeAs~\cite{Watson2018PRB_NaFeAs}. 
However, the spectra also contain bands that are not attributed to the topmost FeAs layer, and it has been a significant challenge to establish which, if any, features can be attributed to the bulk electronic structure of LaFeAsO.  

One of the possible strategies to reveal the real bulk electronic structure of LaFeAsO in ARPES is to deliberately suppress these surface contribution, for example by dosing alkali metals on freshly cleaved surface of the crystal in ultra-high vacuum~\cite{Feng_LaOFeAs,Ding_LaOFeAs}. 
However, this also leads to smearing of the photoemission spectra and therefore significantly reduces the quality of obtained data and confidence in the obtained values of the Fermi surface dimensions. 
Moreover, while single surface terminations can be accessed using microscopic techniques such as scanning tunneling microscopy (STM)~\cite{Wang_STM1111}, the much larger spot-size of conventional ARPES light sources ($\sim$1\ mm for the He-lamp lab-based and $\sim$100\ micron for the synchrotron radiation sources) means that previous studied have all measured the superposition of the signals from two different terminations.

In order to overcome these issues and directly probe the electronic structure of different terminations of LaFeAsO, we used the nano-ARPES technique, in which one could achieve sub-micron spatial resolution by focusing incoming light with Fresnel zone plates (FZP). This method has been successfully applied to other iron-based superconductors to reveal electronic structure of the single antiferromagnetic domain as in case of BaFe$_2$As$_2$~\cite{Matt_NanoARPES}, or nematic domain as in case of FeSe~\cite{Luke_nanoARPES}.
By analysing the chemical shift and overall intensity of As~$3d$ core levels, we identify the spatial distribution of surface terminations, and obtain different valence band spectra from different positions. From this, we can directly assign the bands observed near the Fermi level as originating from either a FeAs termination, or the FeAs layer below an LaO termination. 
Our results are compared with DFT calculations allowing for relaxation at the surface~\cite{eschrig2010calculated}. 

\section{Methods}
High quality single crystals of LaFeAsO were synthesized by solid state single crystal growth method as described elsewhere \cite{KAPPENBERGER_crystalgrowth}.
Nano-ARPES measurements were performed using 70 eV photon energy in linear horizontal polarisation at the nano-ARPES branch of I05-ARPES beamline, using a Scienta DA30 electron analyser. 
Reference high resolution ARPES data at 70 eV photon energy in linear horizontal polarisation were obtained at the HR-ARPES branch of the same I05-ARPES beamline at Diamond Light Source  using R4000 electron energy analyser~\cite{Moritz_RevSciInstr_I05}. 
All measurements were taken at low temperatures (below 40~K for nano-ARPES and below 10~K for HR-ARPES) and in ultra-high vacuum (better than $2\times{}10^{-10}$ mbar). 
Samples were cleaved in vacuum and oriented along the $\Gamma - M$ direction of the Brillouin Zone.
The raw data for the spatial map of As~$3d$ core levels data have been normalized to take account of the MCP detector sensitivity, using reference data measured in swept mode.
DFT calculations were performed using open-source package Quantum Espresso \cite{Giannozzi_2017} using Projected Augmented Wavefunctions (PAW) and the Perdew-Burke-Ernzerhof (PBE) exchange-correlation functional. 
A kinetic energy cutoff of 80 Ry and 960 Ry were used for the wavefunction and charge density respectively for both bulk and slab calculations. A $k$-grid of $8\times{}8\times{}4$ and $6\times{}6\times{}1$ were used for the bulk and slab calculations respectively. 
For the slab calculations, 5 layers of FeAs were interspaced with 4 layers of LaO to create a FeAs terminated surface, and \textit{vice versa} for the LaO terminated surface, equivalent to Ref.~\cite{eschrig2010calculated}. 
Then 15~\AA{} of vacuum was added to both sides, and these structures were then allowed to relax until the total force was less than 0.001 a.u. 
For details, please refer to the input files attached in the supplemental material.

\section{Results}
The crystal structure of LaFeAsO consists of FeAs layers, the essential building blocks of the Fe-based superconductor family, alternated with layers of LaO, which can be considered as an insulating spacer layer. 
In order to prepare a surface for photoemission spectroscopy, the material is cleaved between the layers, leaving two possible surface terminations \cite{eschrig2010calculated}. 
As shown schematically in Fig.~\ref{fig1}(a), both of the terminations result in polar surfaces, with different charge from the bulk value, leading to unique 2D electronic states at each of these surfaces. ARPES is a surface-sensitive technique that can access such surface states, however with a macroscopic beam spot, it is inevitable that the signal from both terminations is superimposed.

\begin{figure}[tb]
\centering
\includegraphics[width=\linewidth]{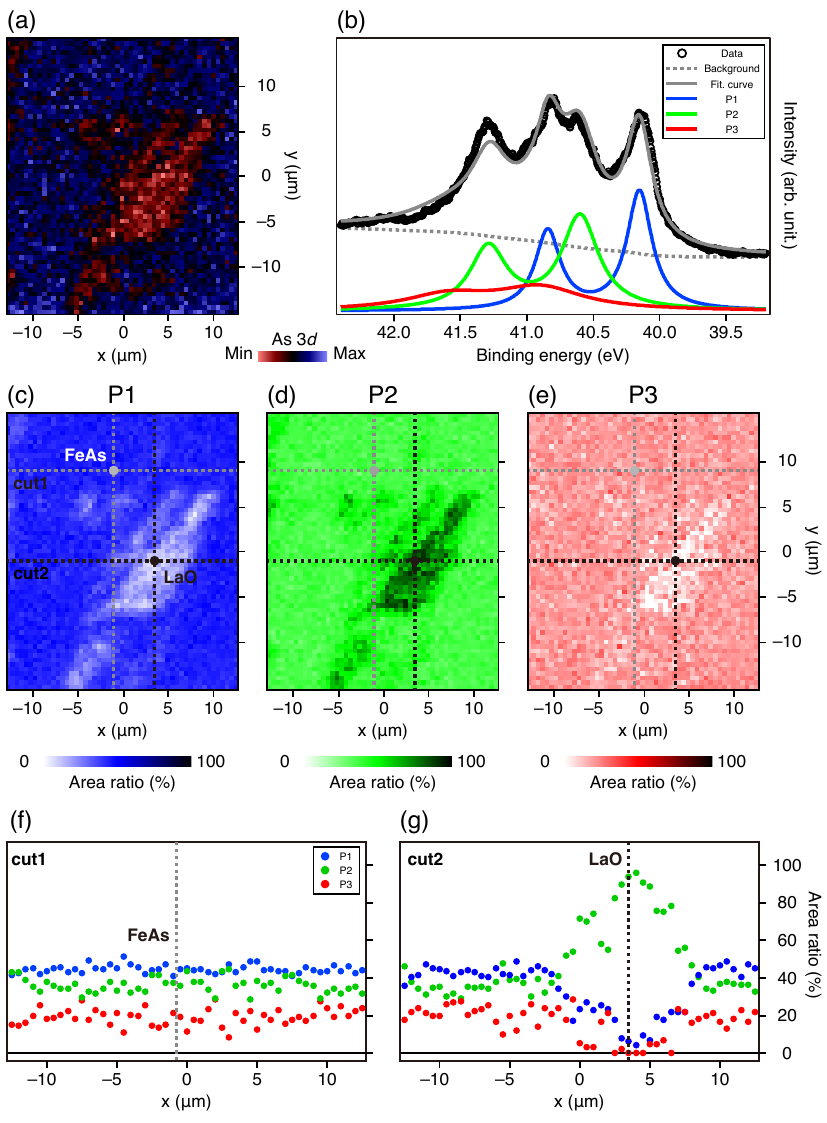}
\caption{\textbf{As~$3d$ spatial maps of LaFeAsO}
(a) spatial resolved 2D map of As~$3d$ core levels total intensity; (b) spatially integrated As~$3d$ core level spectra with fit showing contribution from surface (P1), subsurface (P2) and bulk (P3) FeAs layers;
(c,d,e) spatial resolved 2D maps of As~$3d$ core levels fit contributions from surface, subsurface and bulk FeAs layers; (f,g) area ratio of the fit components for FeAs and LaO terminations.}
\label{fig2}
\end{figure}

To establish the termination-dependent electronic structures, we used a small beam spot ($\sim$800 nm), and scanned the sample while measuring the photoelectrons to map out the sample both spatially and electronically. 
In order to quantitatively map the spatial distribution of terminations, we measured As~$3d$ core-level spectra across a region of the sample. 
The total intensity, shown in Fig.\ref{fig2}(a), is fairly uniform except for a region of lower intensity (red color). 
On the basis of the intensity alone, we could already assign this as a region of LaO termination, since the As atoms in this case are in the subsurface layer and therefore should be weaker in intensity compared to FeAs terminations. 
However the chemical shift analysis of the As~$3d$ core states in the different layers allows a more thorough fingerprinting of these different terminations.

The summation of the As~$3d$ spectra across the whole spatial map is presented in Fig.~\ref{fig2}(b), clearly showing contributions from several peaks, identified in fitting analysis as doublets P1, P2 and P3. 
While the spectrum is complex, we can identify the doublet at lowest binding energy, P1, as arising from the topmost As on an FeAs termination, since it has the largest chemical shift. 
This assignment is further supported by our additional photon-energy and geometry-dependent measurements performed with a larger beam spot on mixed-termination samples (see Supplementary Figure~\ref{figS1}) which show that the P1 peak is most prominent in conditions which are most surface-sensitive, namely at lower photon energy and at grazing incidence. 
Our assignment of P1 is also consistent with previous experiments where the intensity of low binding energy peak drastically decreased with alkali metal dosing of the surface~\cite{Feng_LaOFeAs,Ding_LaOFeAs}.
The doublet with highest binding energy, P3, shows the opposite dependence to P1, and thus corresponds to bulk-like As layers from deeper in the sample, where the surface charge is fully screened. 
The doublet with intermediate binding energy, P2, can be attributed to the As atoms that are neither bulk-like nor on the topmost surface; either from the As atoms below the Fe within the FeAs surface termination, or from the top As in the sub-surface FeAs layer below an LaO termination. 

Using such three doublet fit of the summed data to constrain parameters, we then fit the spectra on a pixel-by-pixel basis, allowing the amplitude of the peaks to vary, leading to the spatial maps displaying the dependence of the intensity of the P1, P2 and P3 peaks in Fig.~\ref{fig2}(c-e), from which we also take two line cuts in (f,g). 
For cut~1 in Fig.~\ref{fig2}(f), corresponding to a uniform region of FeAs termination, all the peaks P1, P2 and P3 contribute to the spectra, with very small variation. 
However, for cut~2 in Fig.~\ref{fig2}(g), we find a region approximately 10 microns wide, in which the P2 peak completely dominates the spectral intensity, while P1 and P3 are suppressed. 
The suppression of the P1 peak, which was assigned to the FeAs termination, implies that the dominant P2 peak here comes from the LaO termination (i.e. from the top As in the sub-surface FeAs layer). 
Thus, our spatial mapping analysis not only identifies the different contributions to the As~$3d$ core level spectra, but also allows us to confidently place the beam either on a region with FeAs termination, or to measure on the areas with LaO termination. 

\begin{figure}[b]
\centering
\includegraphics[width=\linewidth]{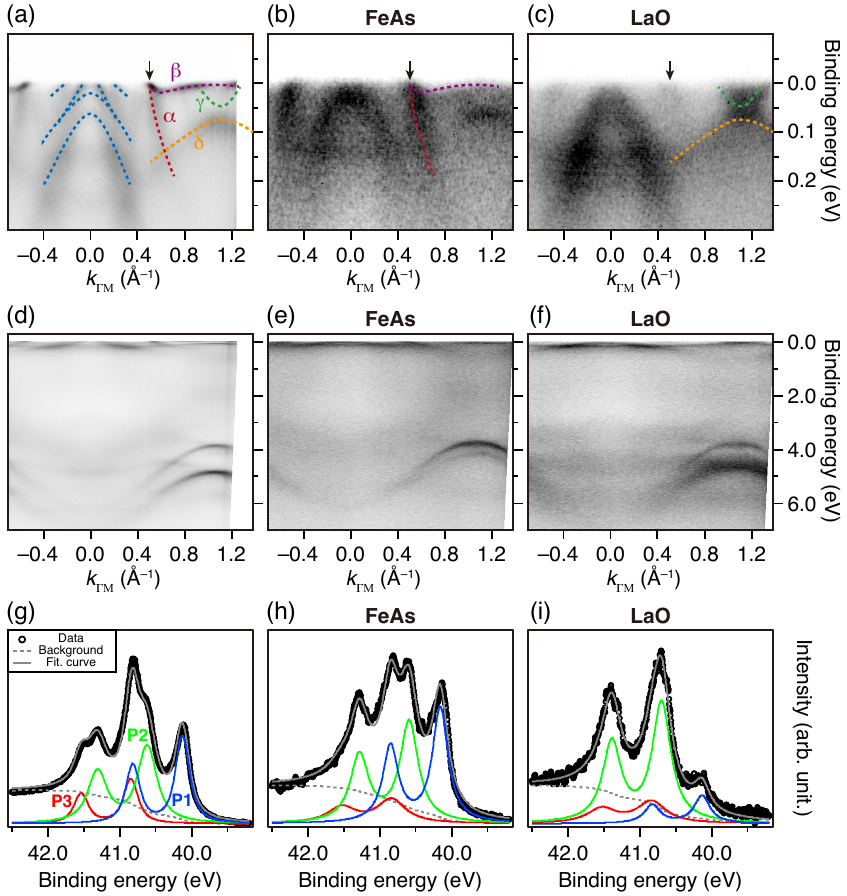}
\caption{\textbf{Termination dependent electronic structure of LaFeAsO}
(a,b,c) band dispersions at the Fermi level from large beam spot at high-resolution branch and FeAs and LaO terminations submicron beam spot at nano-ARPES branch;
(d,e,f) valence band dispersions from large beam spot at high-resolution branch and FeAs and LaO terminations submicron beam spot at nano-ARPES branch;
(g,h,i) As~$3d$ core level spectra from large beam spot at high-resolution branch and FeAs and LaO terminations submicron beam spot at nano-ARPES branch.}
\label{fig3}
\end{figure}

With the knowledge of the distribution of the terminations established, we turn to the corresponding electronic structure of LaFeAsO. 
First, in Fig.~\ref{fig3}(a), we show the measured band dispersions at the Fermi level and valence bands using high-resolution ARPES, but with a beam spot of $\sim$80 $\mu$m. 
From the conventional ARPES data of LaFeAsO in Fig.~\ref{fig3}(a), we label a number of the most easily identifiable bands as $\alpha$,$\beta$,$\gamma$,$\delta$; these have all been reported in previous studies~\cite{KaminskiLaFeAsO,Feng_LaOFeAs,Ding_LaOFeAs}. 
While there have been several attempts to identify the origin of all the bands from measurements with a macroscopic beam spot \cite{Ding_LaOFeAs} and comparison with DFT calculations~\cite{eschrig2010calculated}, our methodology relies on more direct approach to measure electronic structure of various surface terminations. 

In Fig.~\ref{fig3}(b), taken on a FeAs termination, we observe most prominently the $\alpha$ band, which confirms it as a surface FeAs band \cite{KaminskiLaFeAsO,Feng_LaOFeAs,Ding_LaOFeAs}, which is the brightest feature in most measurement conditions. 
The kink-like structure of the $\alpha$ band has been previously discussed \cite{Liu_2010_CeFeAsO_alphakink,Feng_LaOFeAs} and is evidence of strong electron-electron interactions in the hole-doped surface layer. 
In addition, the $\beta$ band contributes to the spectrum on the FeAs termination. 
On the other hand, these features are all suppressed when the spot is moved onto the LaO termination, shown in Fig.~\ref{fig3}(c), and instead the spectrum includes an electron-like $\gamma$ band at the M point. 
This electron-like band, and also the hole-like $\delta$ band, originate from  Fe-derived bands from the subsurface FeAs layer.

The electronic structure that emerges from the FeAs subsurface layer below the LaO termination has previously been obscured in measurements on mixed termination samples, since ARPES is highly surface sensitive and the bands from the FeAs termination dominate the intensity. 
However, it is noteworthy for several reasons. 
First, the $\gamma$ electron band - most likely comprising two electron-like bands in close proximity, unresolved in the current measurements - is reasonably large, with a $k_F\approx{}$~0.25~\AA{}$^{-1}$ and $E_B\approx{}$-80~meV and the observed hole-like states at $\Gamma$ in Fig.~\ref{fig3}(c) are fully occupied, implying an uncompensated system with net electron doping. 
The subsurface layer was predicted to have a net charge of 0.114 electrons per Fe compared with the bulk material \cite{eschrig2010calculated}, qualitatively consistent with our experiments. 
Conceptually, the LaO layer does a relatively poor job of screening the polar surface charge, such that there is still a substantial additional charge in the subsurface layer compared with a bulk FeAs layer. 
Second, we do not observe any electronic reconstruction, which implies that the effective electron doping of the subsurface FeAs layer is sufficient to move beyond the antiferromagnetic phase, which in the bulk survives to $x\approx{}$0.05~\cite{Scaravaggi_PRB_2021-LaFeCoAsO}.
Third, we note an fascinatingly close analogy between the electronic structure of the subsurface FeAs and the high-$T_c$ superconducting monolayer FeSe/STO: both are strictly 2D systems, involving an interface with an oxide material, and in which a net electron doping of $\sim${}0.1 electrons per Fe occurs to yield a Fermi surface containing only electron pockets~\cite{Tan_NatMatt_-2013-FeSe-STO_SC,He_PNAS_2014-FeSe-STO}.

\begin{figure}[b]
\centering
\includegraphics[width=\linewidth]{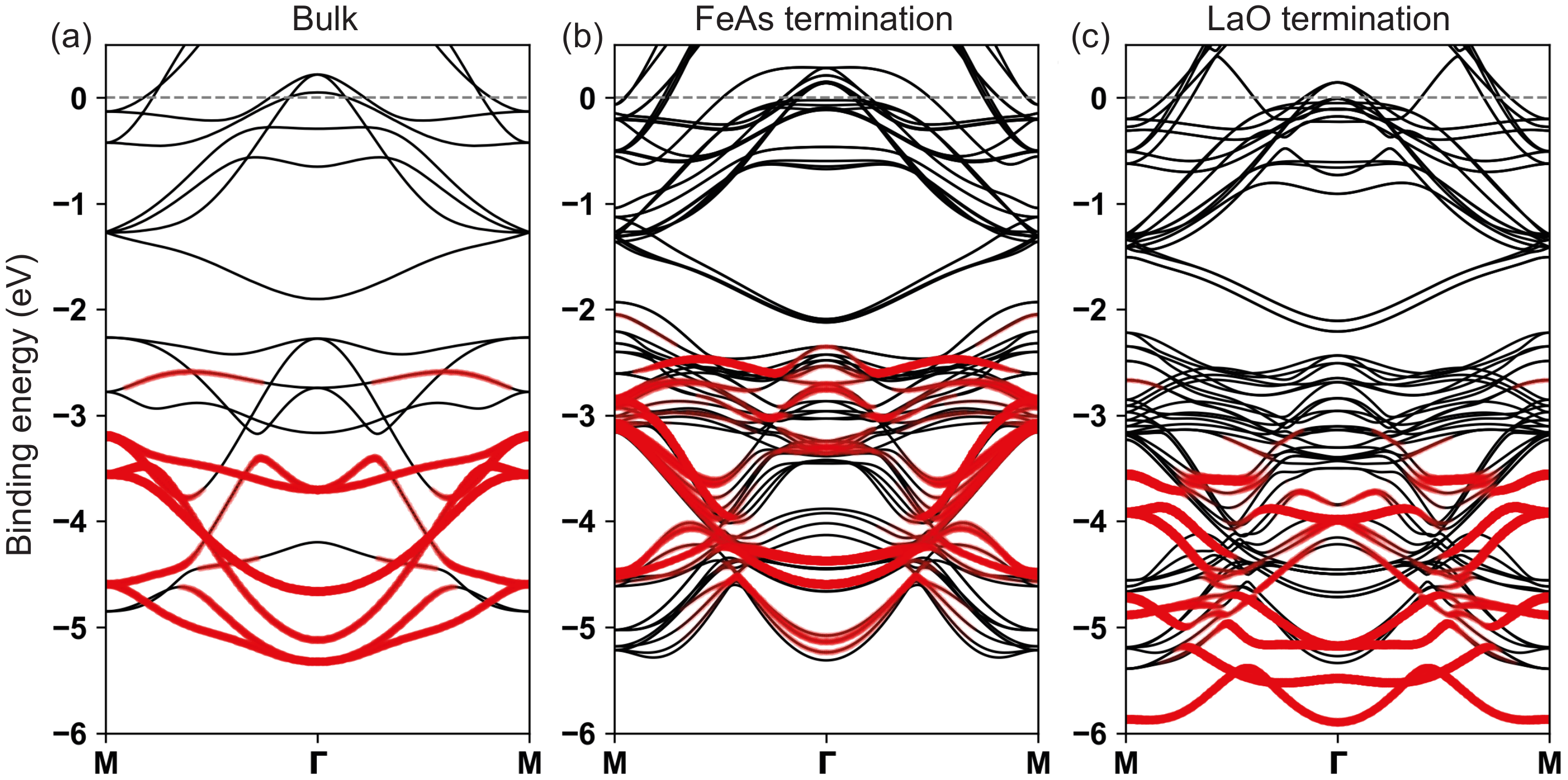}
\caption{\textbf{DFT electronic structure of LaFeAsO}
- simulations of the band structure highlighting the oxygen (red) dominated valence bands for a) a bulk paramagnetic crystal, b) an FeAs terminated surface, c) a LaO terminated surface. In (b,c) we only colour  the oxygen weight arising from the closest layer to the surface.
}
\label{fig4}
\end{figure}

Finally, we turn our attention to the valence bands at 4\textendash6~eV binding energy, which provide a complementary understanding of the observed surface terminations.
In Fig.~\ref{fig3}(d), we present high resolution measurements of the valence band using an $\sim${}80 $\mu$m beamspot, observing two bright and sharp bands around the M point, connected to band minima with weaker intensity at the $\Gamma$ point. 
By comparing these dispersions with DFT-based calculations, as shown in Fig.~\ref{fig4}(a-c), we assign these as O~$2p$ bands. Although As~$4p$ states are also expected at these binding energies, those bands tend to be only weakly observed in photoemission measurements due to high-energy interactions in the FeAs layer \cite{Evtushinsky2017PRB}. 

When we look at the distinct valence band data from the two terminations, we observe that the LaO termination (Fig.~\ref{fig3}(e)) has the O $2p$-derived valence bands at generally higher binding energy than the FeAs termination (Fig.~\ref{fig3}(f)). 
This can be naturally explained, since the heavy electron doping of the LaO layer pushes the surface layer oxygen states to deeper binding energies. 
The overall downward shift of the oxygen bands on the LaO surface is further reproduced by DFT slab calculations on FeAs and LaO-terminated surfaces, shown in Fig.~\ref{fig4}(b) and (c). 
This shift independently confirms the heavily electron doped nature of the LaO surface state. 
Meanwhile, the calculations in Fig.~\ref{fig4}(b) show that the oxygen states in the subsurface LaO layer beneath the FeAs termination have dispersions and binding energies much closer to the bulk ones, since the FeAs layer screens the surface charge quite effectively. 

\section{Discussion}

In this work, using nano-ARPES as a local probe method, we established a strong dependence of the electronic structure with surface termination due to significant charge transfer at the polar surface of the LaFeAsO crystal.
Because of this, it is not possible to examine the bulk electronic structure using surface sensitive techniques like STM and ARPES. 
On the contrary, we can only observe two very different surface electronic structures with opposite hole or electron doping. 
Our results are in good qualitative agreement with termination dependent STM data~\cite{Zhou_PRL}.

More importantly, it explains, why unlike in non-polar pnictides ~\cite{NaFeAs_AFMSDW,Feng_NaFeAs,MYi_NaFeAs}, the band folding corresponding to the antiferromagnetic (AFM) order below $T_N$ $\sim${}135~K was not been detected in LaFeAsO by ARPES. 
Although we were able to separately measure the electronic structure of the FeAs surface termination and the FeAs subsurface under an LaO termination, neither of these are truly representative of the bulk electronic structure of LaFeAsO.
Moreover, direct photoemission from deeper layers is suppressed beyond detection due to the short escape depth of low energy photoelectrons. Therefore we conclude that the true bulk electronic structure of LaFeAsO, including the expected electronic reconstruction in the AFM phase, cannot be accessed by ARPES, at least in the VUV photon energy range. 
Instead, based on this study, we can interpret with confidence the previous ARPES data for LaFeAsO obtained with conventional (large) beam spot~\cite{Shen_LaOFeAs,KaminskiLaFeAsO,Feng_LaOFeAs,Ding_LaOFeAs,CKim2020_LaFeAsO} as a superposition of signals from two different surface terminations.    

As we have shown, the charge transfer induced surface doping is large enough to suppress the AFM phase for the parent compound. 
So one could ask the question, what are the implications of this effect on the superconductivity in the ``1111” oxopnictide family?
Despite their surface-sensitivity, both STM~\cite{Fasano_PRL-STM} and ARPES~\cite{Liu-arxiv_0806_2147-ARPES} have reported an opening of the superconducting gap in fluorine doped single crystals. 
While STM measures the spectral function integrated along all pockets of the Fermi surface, ARPES provides direct information on the momentum dependence of the superconducting order parameter. 
A detailed momentum dependent ARPES study~\cite{Kondo_PRL-ARPES} revealed an almost isotropic superconducting gap for the large hole-like band around the $\Gamma$-point.
It is clear from our results, this particular Fermi surface belongs to the FeAs termination. 
Because FeAs termination electronic structure is strongly holed doped, it is unlikely that this top layer is superconducting by itself. 
A much more plausible scenario is that surface state superconductivity is induced by proximity to the the superconducting bulk crystal, an intriguing prospect which merits further investigation.

There are numerous analogies between our work on LaFeAsO, and recent work on layered delafossite materials such as PdCoO$_2$, where there has been interest in the novel electronic states that appear at the polar surfaces after cleaving~\cite{Phil_PdCoO2}.
In fact, for any `naturally heterostructured' material consisting of alternating layers, the problem of multiple possible surface terminations contributing to ARPES spectra with a conventional beam spot is widespread. 
However, the nano-ARPES methodology that we use here, and in other recent experiments~\cite{Phil_PdCoO2, Iwasawa_PRB_2018, Iwasawa_PRB_2019} opens the door to obtaining single-termination spectra. 

\section{Conclusions}
Using angular-resolved photoemission spectroscopy with a sub-micron focused beam spot (nano-ARPES) we measured the electronic structures for both LaO and FeAs terminations in LaFeAsO. 
While large bulk of the ARPES studies is concentrated on a narrow energy window around the Fermi level, we have shown that the detailed combined analysis including the nano-ARPES photoemission signal from much higher binding energies allows for a direct decomposition of the observed intensity into the contributions from different surface layers. 
Our data shows different Fermi surfaces, valence band electronic structure and core-level spectra for different surface terminations. 
Our experiments confirm the theoretical calculations and explain controversy of the previous conventional ARPES measurements. 
These results give direct evidence for the surface-driven electronic structure reconstruction in LaFeAsO, including formation of the termination-dependent surface states at the Fermi level.
The nano-ARPES technique used to study this prototypical ``1111" compound, has and can be successfully applied to study different polar materials with different surface terminations.

\begin{acknowledgments}
\section{Acknowledgments}
We thank Diamond Light Source for access to beamline I05 (proposal numbers SI15074 and SI23890) that contributed to the results presented here. 
LCR acknowledges funding from the Royal Commission for the Exhibition of 1851.
The work at IFW was supported by the Deutsche Forschungsgemeinschaft (DFG) through the Priority Programme SPP1458. 
SA thanks the DFG for funding (AS 523\textbackslash{}4-1 \& 523\textbackslash{}3-1).
\end{acknowledgments}

\bibliography{LaFeAsO_nanoARPES.bib}

\clearpage
\appendix
\counterwithin{figure}{section}
\counterwithin{table}{section}
\renewcommand{\thefigure}{A\arabic{figure}}
\renewcommand{\thetable}{A\arabic{table}}
\setcounter{figure}{0}
\setcounter{table}{0}

\section{Appendix - Reference core level data}
High energy resolution LaFeAsO As~$3d$ reference core level spectra were measured at HR-ARPES end-station with 80 micron beam spot size.
\begin{figure}[hbt!]
\centering
\includegraphics[width=\linewidth]{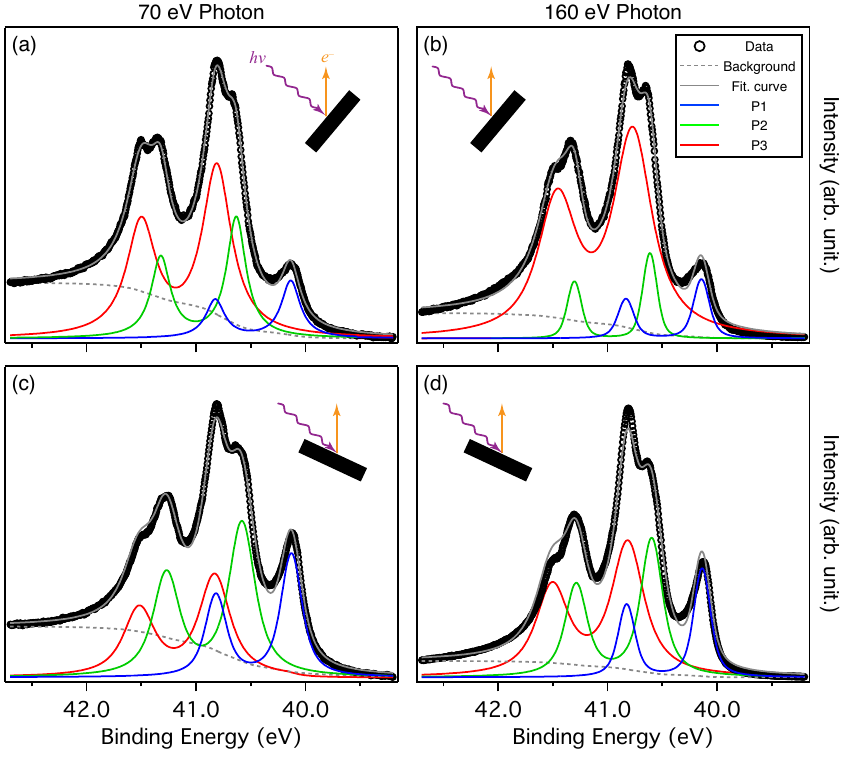}
\caption{\textbf{LaFeAsO As~$3d$ XPS:}
(a) photon energy 70 eV and normal incidence, (b) photon energy 160 eV and normal incidence, (c) photon energy 70 eV and grazing incidence, (d) photon energy 160 eV and grazing incidence. 
}
\label{figS1}
\end{figure}

Corresponding As~$3d$ reference core level spectra were fitted with 3 sets of doublets P1, P2 and P3.
\begin{table}[hbt!]
\begin{ruledtabular}
\begin{tabular}{cccccccc}

   & (a) 70eV NI & (b) 160eV NI & (c) 70eV GI & (d) 160eV GI \\
   \colrule
P1 & 0.128       & 0.084        & 0.257       & 0.183        \\
P2 & 0.276       & 0.103        & 0.416       & 0.327        \\
P3 & 0.596       & 0.813        & 0.327       & 0.490       
\end{tabular}
\caption{LaFeAsO As~$3d$ XPS fit results}
\label{TableS1}
\end{ruledtabular}
\end{table}
\end{document}